\documentclass[conference]{IEEEtran}
\IEEEoverridecommandlockouts
\usepackage{cite}
\usepackage{amsmath,amssymb,amsfonts}
\usepackage{algorithmic}
\usepackage{graphicx}
\usepackage{textcomp}
\usepackage{xcolor}
\usepackage{hyperref}
\usepackage{amsthm}
\theoremstyle{definition}
\newtheorem{definition}{Definition}
\usepackage{enumitem}

\def\BibTeX{{\rm B\kern-.05em{\sc i\kern-.025em b}\kern-.08em
    T\kern-.1667em\lower.7ex\hbox{E}\kern-.125emX}}
\begin{document}
\newcommand{\ECDSA}{{\sf ECDSA}}
\newcommand{\Setup}{{\sf Setup}}
\newcommand{\KeyGenS}{{\sf KeyGen}_{\sf S}}
\newcommand{\KeyGenN}{{\sf KeyGen}_{\sf N}}
\newcommand{\Sign}{{\sf Sign}}
\newcommand{\Receive}{{\sf Receive}}
\newcommand{\Convert}{{\sf Convert}}
\newcommand{\TkVerify}{{\sf TkVerify}}
\newcommand{\Comfirm}{{\sf Comfirm}}
\newcommand{\Disavow}{{\sf Disavow}}
\newcommand{\Valid}{{\sf Valid}}
\newcommand{\pkS}{{\sf pk}_{\sf S}}
\newcommand{\skS}{{\sf sk}_{\sf S}}
\newcommand{\pkN}{{\sf pk}_{\sf N}}
\newcommand{\skN}{{\sf sk}_{\sf N}}
\newcommand{\pk}{{\sf pk}}
\newcommand{\sk}{{\sf sk}}
\newcommand{\vk}{{\sf vk}}
\newcommand{\tk}{{\sf tk}}
\newcommand{\Verify}{{\sf Verify}}
\newcommand{\KeyGen}{{\sf KeyGen}}
\newcommand{\accept}{{\sf accept}}
\newcommand{\reject}{{\sf reject}}
\newcommand{\deposit}{{\sf deposit}}
\newcommand{\parameter}{{\sf par}}
\newcommand{\ZKPK}{{\rm ZKPK}}
\newcommand{\VC}{{\sf V}_{\sf C}}
\newcommand{\VD}{{\sf V}_{\sf D}}
\newcommand{\G}{\mathbb G}
\newcommand{\FS}{{\it F_S}}
\newcommand{\FN}{{\it F_N}}
\newcommand{\MS}{{\it M_S}}
\newcommand{\MN}{{\it M_N}}
\newcommand{\Nominee}{{\rm Nominee}}

\def\qed{\hfill $\Box$}

\newtheorem{theorem}{Theorem}[section]
\newtheorem{Definition}{Definition}[section]

\title{A Smart Contract-based Non-Transferable Signature Verification System using Nominative Signatures
}

\author{
\IEEEauthorblockN{Hinata Nishino}
\IEEEauthorblockA{\textit{Kanazawa University, Japan}}\\
\and
\IEEEauthorblockN{Kazumasa Omote}
\IEEEauthorblockA{\textit{University of Tsukuba, Japan}}\\
\and
\IEEEauthorblockN{Keita Emura}
\IEEEauthorblockA{\textit{Kanazawa University, Japan}}
\IEEEauthorblockA{\textit{AIST, Japan}}\\
}

\maketitle

\begin{abstract}
Nominative signatures allow us to indicate who can verify a signature, and they can be employed to construct a non-transferable signature verification system that prevents the signature verification by a third party in unexpected situations. For example, this system can prevent IOU/loan certificate verification in unexpected situations. 
However, nominative signatures themselves do not allow the verifier to check whether the funds will be transferred in the future or have been transferred.
It would be desirable to verify the fact simultaneously when the system involves a certain money transfer such as cryptocurrencies/cryptoassets.  
In this paper, we propose a smart contract-based non-transferable signature verification system using nominative signatures. 
We pay attention to the fact that the invisibility, which is a security requirement to be held for nominative signatures, allows us to publish nominative signatures on the blockchain. Our system can verify whether a money transfer actually will take place, in addition to indicating who can verify a signature. We transform the Hanaoka-Schuldt nominative signature scheme (ACNS 2011, IEICE Trans. 2016) which is constructed over a symmetric pairing to a scheme constructed over an asymmetric pairing, and evaluate the gas cost when a smart contract runs the verification algorithm of the modified Hanaoka-Schuldt nominative signature scheme.
\end{abstract}

\begin{IEEEkeywords}
Smart contract, Nominative signatures, Non-transferability
\end{IEEEkeywords}

\section{Introduction}

\subsection{Background} 

There are many situations where it is necessary to verify who has issued certain information. Generally, using a digital signature scheme makes it possible to verify the issuer of the information. However, since the usual digital signature scheme allows for public verification, there is a possibility that the signature verification could be executed in unexpected situations for the person handling the information. For example, even information that one does not want to disclose to third parties, such as debts, could be verified for its validity.

\begin{figure}[h]
\centering
\includegraphics[width=8.5cm]{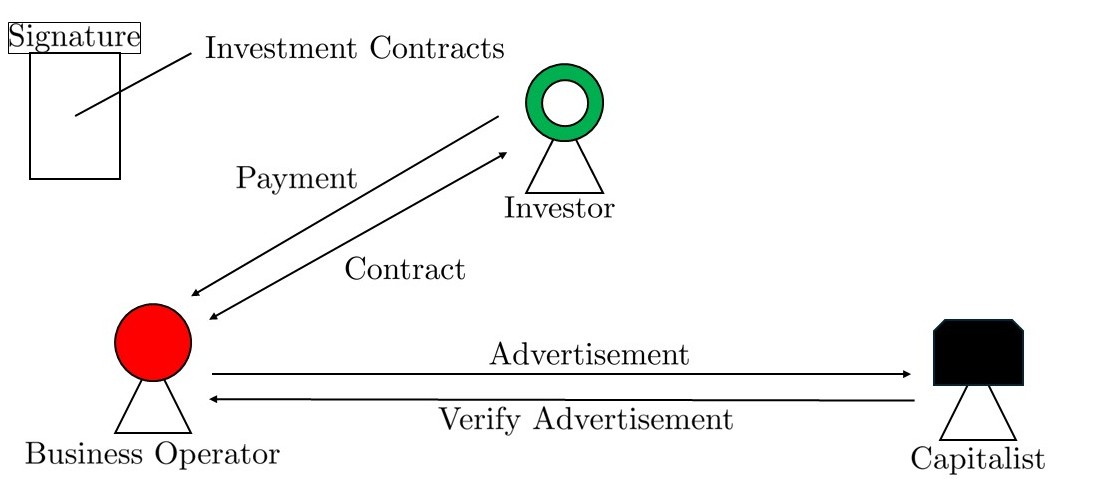}
\caption{Advertisement of Investment Contracts}
\label{fig:chara}
\end{figure}

Here, we introduce an advertisement of investment contracts as a specific example where the public verifiability of signatures becomes a problem (See Fig~\ref{fig:chara}). Assume that investment contracts are made between a business operator who conducts business and an investor who makes investments. When a business operator seeks to receive more investments, it is effective to appeal to other capitalists that they have received investments from investors. When a business operator appeals that they have received investments, it is assumed that the business operator produces and publishes a signature on the contract so that the information can be verified by third-party capitalists as being issued by the business operator.

Specifically, by attaching a signature to a contract indicating that the business operator and the investor have entered into a financial contract, it serves as evidence of the contract between the business operator and the investor. 
When the business operator uses the contract to advertise to capitalists, the capitalists can indeed verify that the information was issued by the business operator.%
\footnote{Strictly speaking, it is necessary to separately verify that the signature verification key belongs to the business operator using PKI.}
At the first sight, the system seems to be feasible when the investor will complete the financial assistance. However, due to the public verifiability, anyone, who obtains the business operator's verification key and the signed contract, can verify the signature. This raises concerns that capitalists could use this investment information although the business operator does not know this fact. A capitalist could cause trouble by proving the validity of the contract to a third party without the business operator's and investor's awareness. For example, a capitalist might commit investment fraud based on the investment information.
Therefore, it is necessary to 
\begin{quote}
\emph{appropriately control who can verify the signature}.  
\end{quote}
\noindent 
Additionally, the investor has an incentive to attract more investments to the business operator they are investing in, as it increases the likelihood of the business's success. Note that considering the possibility that the business operator might solicit funds through false advertising, the verifiability of the business operator's investment should also involve the investor. 

From the perspective of the capitalist, it is desirable to decide whether the capitalist investments or not after confirming the investor has actually invested. 
It might be possible to verify this fact by some means after funds have actually been transferred from the investor to the business operator. 
However, a certain time lag is expected between the conclusion of the contract and the transfer of funds by considering the time required for the investor to prepare the funds. 
It is unreasonable to wait for the transfer before starting the advertisement when the business operator wants to advertise to attract further investments by using the fact of the transfer from the investor. 
Furthermore, since the signature is independent to the funds,  and the transfer is conducted between the business operator and the investor, the capitalist cannot confirm the presence or absence of the transfer through signature verification. 
Therefore, even if the investor will not transfer the funds contrary to the contract, the capitalist cannot verify this. This could result in the capitalist bearing the risk unilaterally. To solve these problems, a method that can prove that 
\begin{quote}
\emph{the funds will be transferred in the future, even before the transfer,} 
\end{quote}
\noindent 
is necessary. The usual publicly verifiable signature scheme does not meet these requirements.

\subsection{A Naive Solution and Its Limitation}

As a naive and simple solution, we consider to employ nominative signatures~\cite{KPW96}. In a nominative signature scheme, a new entity called Nominee is defined. A signer and a nominee jointly generate a signature which is called a nominative signature, and the nominee proves the validity of the nominative signature through an interactive protocol with the verifier. Without the nominee, even the signer or a verifier who has once participated in verification cannot prove the verification result to a third party. 
By taking advantage of this property, we can expect to construct a system in which the business operator and the investor can control the verifiability of signed contracts. Specifically, the business operator is assigned as a signer and the investor as a nominee, and they jointly generate a nominative signature (on a contract). The business operator asks the investor to prove the validity of the contract to the capitalist, and the investor communicates with the capitalist to prove it. Due to the property of nominative signatures, the capitalist cannot prove the validity of the contract to a third party. The above system allows us to appropriately control who can verify the signatures. Note that nominative signatures themselves are still independent to the funds provided. Due to this reason, nominative signatures are not effective to prove that the funds will be transferred in the future before the funds are transferred.

\subsection{Our Contribution}
In this paper, we propose a non-transferable signature verification system. 

\begin{enumerate}
    \item In addition to a nominative signature scheme, we employ smart contracts to connect a signature with money transfers. 
    \begin{itemize}
        \item Concretely, we connect a signature with money transfers by the following procedure: A business operator and an investor jointly generate a nominative signature on \emph{the program source code} of the smart contract. 
    \end{itemize}
    \item We also evaluated the performance of the proposed system using pre-compiled contracts provided by smlXL.inc~\cite{smlXL}. 
    \begin{itemize}
        \item We employ the pairing-based Hanaoka-Schuldt nominative signature scheme~\cite{HanaokaS16} in our evaluation. 
    \end{itemize}
\end{enumerate}

\noindent 
Smart contracts allow various processes to be executed according to pre-defined and publicly disclosed contracts. Therefore, it is expected that the system will be configured in such a way that the capitalist can verify the transfer as well. The proposed system is briefly explained as follows. 

\begin{enumerate}
    \item We introduce a smart contract that manages a transfer. 
    \item A business operator and an investor jointly generate a nominative signature on the program source code of the smart contract, and store the nominative signature on the smart contract. 
    \item An operation using a nominative signature in the smart contract is run after the investor is ready to transfer, which triggers the transfer. 
    \end{enumerate}
A wallet on the blockchain can be viewed by anyone, so the capitalist can check whether or not the transfer has been made. Also, the capitalist can check whether the funds will be transferred in the future or not (by the trigger described in the program), even before the transfer, by checking the program source code of the smart contract. 
Note that it is easy to verify whether or not the transfer to the business operator has been done by checking the transaction after the transfer, and that the investor cannot illegally withdraw the funds from the business operator's wallet since the funds are locked by the wallet. In the proposed system, a nominative signature is stored on a smart contract and disclosed to the public. However, no information of the business operator and the investor is revealed due to the invisibility of the underlying nominative signature scheme. 
Moreover, due to the security of the underlying nominative signature scheme, it is guarantees that both the business operator and the investor agree on the creation of the nominative signature. It also employed to prevent the business operator to run the smart contract without the investor's approval, and to prevent the investor from proving to a capitalist that they are willing to make an investment. 

\medskip
\noindent\textbf{Differences from the proceedings version}. An extended abstract appeared at the 20th Asia Joint Conference on Information Security (AsiaJCIS) 2025~\cite{NishinoOE25}. This is the full version. We declare that all figures presented in this full version are identical to those published in the proceedings version. 

As the additional content, given in Section~\ref{NSoverAP}, we transform the Hanaoka-Schuldt nominative signature scheme~\cite{HanaokaS16} which is constructed over a symmetric pairing to a scheme constructed over an asymmetric pairing, and evaluate the gas cost when a smart contract runs the verification algorithm of the modified Hanaoka-Schuldt nominative signature scheme. The reason behind the transformation is explained as follows. 

\begin{itemize}
    \item A pre-compiled contract smlXL.Inc~\cite{smlXL} that we employed provides elliptic curve operations and pairing computation on Barreto-Naehrig (BN) curves~\cite{BarretoN05} (bn128), i.e., asymmetric pairings. However, the original Hanaoka-Schuldt nominative signature scheme, that was employed as the underlying nominative signature scheme in the proceedings version~\cite{NishinoOE25}, is constructed over symmetric pairings.
    \item We note that asymmetric pairings provide more efficient implementation compared to that of symmetric pairings~\cite{GalbraithPS08}. This is due to the difference in the embedding degree.%
\footnote{Intuitively, the ratio of the size of a point on an elliptic curve (source group) to the size of the finite field (target group). 
For example, for a BN curve with an embedding degree of 12, it is sufficient to set the size of a source group element is 256 bits (then, the size of a target group element is 3,000 bits that guarantees the hardness of the discrete logarithm problem over the target group) whereas a symmetric pairing has an embedding degree of 2, it requires that the size of a source group element is 1,500 bits.} 
On the other hand, the number of pairing operations may increase because some elements may be duplicated due to asymmetric pairings.  \end{itemize}
\noindent 
Because the number of pairing operations is dominant of the gas cost, we transform the Hanaoka-Schuldt nominative signature scheme to a scheme constructed over an asymmetric pairing, and evaluate the gas cost when a smart contract runs the verification algorithm of the modified Hanaoka-Schuldt nominative signature scheme. 
As the result, we confirm that the number of pairing computations in the $\TkVerify$ algorithm (that is run on the smart contract) does not increase compared to that of the original Hanaoka-Schuldt nominative signature scheme.

\subsection{Related Work}

\noindent\textbf{Advanced Cryptography in Blockchain}. Hash functions (such as SHA-256) and digital signatures (such as ECDSA) are widely employed in the blockchain. Towards such a relatively simple cryptographic primitive, some advanced cryptographic primitives have been considered for blockchain-oriented applications: zero-knowledge proofs (e.g.,~\cite{OzdemirB22,ParnoHG016,BunzBBPWM18}), linkable ring signatures (e.g.,~\cite{LiuWW04,LiuW05}), accountable ring signatures (e.g.,~\cite{ChinEO25,BootleCCGGP15}), aggregate signatures (e.g.,~\cite{BonehGLS03}), adaptor signatures (e.g.,~\cite{P17,GerhartSST24,DaiOY22,VanjaniST24,0001BBKM19,MadathilTVFMM23,QinPMSESELYY23}) and so on. 
To the best of our knowledge, no attempt to employ a signature scheme with controllable verifiability (listed below) to smart contracts has been considered so far. Since values stored on the blockchain are made public, providing controllable verifiability of signatures seems to be effective in preserving security or privacy in the blockchain environment.  

\medskip
\noindent\textbf{Signature Schemes with Controllable Verifiability}. 
In addition to nominative signatures, many other signature schemes with controllable verifiability have been proposed. Undeniable signatures~\cite{ChaumA89} require that the verifier needs to run an interactive protocol with the signer, and it can prevent signatures from being verified without the signer's knowledge or consent. Furthermore, the signer can claim that they produced a signature, but cannot claim that they did not produce a signature when they have produced the signature. 
In some cases, it is desirable to be able to use an undeniable signature together with a conventional (i.e., publicly verifiable) signature. Therefore, a convertible undeniable signature scheme~\cite{BoyarCDP90} has also been proposed where the signer can convert a previously issued undeniable signature to a publicly verifiable signature.
In undeniable signatures, the signer is required to be always involved to the verification process that increases the workload of the signer. 
To solve this problem, confirmer signatures have been proposed~\cite{Chaum94} that  introduce a third entity called a confirmer who runs the interactive verification protocol with the verifier. 
Online untransferable signatures~\cite{LiskovM08} have also been proposed as a method to prevent a third party from verifying the validity/invalidity of a signature by observing the interactive protocol run between a signer and a verifier online. In designated verifier signatures\cite{JakobssonSI96}, the signer designates a verifier, and only the designated verifier can verify the signature, and the signer is not involved in the signature verification process itself. Furthermore, the designated verifier cannot convince a third party of the validity/invalidity of the signature.  

As a kind of these signatures with controllable verifiability, nominative signatures have been proposed~\cite{KPW96}. In nominative signatures, which are the dual relationship with undeniable signatures, the signature holder called nominee can prove the validity/non-authenticity of the signature to a third party. Several nominative signature schemes have been proposed so far~\cite{HuangLW08,LiuWHWHMS07,ZhaoY09,HuangW04,LiuCWM07}. 
Schuldt and Hanaoka~\cite{SchuldtH11} formalized a security definition of nominative signatures (we mainly refer to the full version~\cite{HanaokaS16}). 
As an application of nominative signatures, 
a privacy-enhanced access log management mechanism in single-sign on (SSO) systems has been proposed~\cite{NakagawaEHKNOS14,NakagawaNOEHSK17}. The system employs a nominative signature as an access log stored on the system. Due to the invisibility of the underlying nominative signature scheme, no information of access user is reveled from the log whereas users can prove that they have accessed the system. The proposed system is inspired by the SSO system because the invisibility is attractive to preserve privacy in blockchain, especially in a public blockchain where anyone can observe information stored on the blockchain.

\section{Roles of ECDSA in Ethereum}

In Ethereum, ECDSA signatures are required for the transfer of cryptocurrencies/cryptoassets. 
Note that the underlying ECDSA signature scheme in Ethereum is not the usual one and is called recoverable ECDSA in Ethereum Yellow Paper~\cite{ETHYellow} where it provides the key recovery property: the verification key is recovered from a signature and a message. The following is a brief overview of the ECDSA signature verification process in Ethereum. 
There are two entities: a sender and a receiver of the funds. The sender generates an ECDSA signature on a transaction $M$ using own secret signing key where (hash value of) the public verification key is the address of its wallet, and sends the transaction with the ECDSA signature to the receiver. The receiver recovers the verification key from the signature and the message. If (the hash value of) the recovered verification key matches the sender's address, it is accepted as a valid signature. 

In the proposed system, ECDSA signatures are generated when funds are transferred from an investor's wallet. 
To avoid any confusion, we do not explicitly specify the recovery phase in the proposed system and employ the following syntax. 
Let ($\ECDSA.\KeyGen$, $\ECDSA.\Sign$, $\ECDSA.\Verify$) be the ECDSA scheme. The key generation algorithm is denoted as $(\vk, \sk) \leftarrow \ECDSA.\KeyGen(1^{\lambda})$ where $\lambda\in\mathbb{N}$ is a security parameter, $\vk$ is a verification key, and $\sk$ is a signing key. The signing algorithm is denoted as $\sigma_{\ECDSA} \leftarrow \ECDSA.\Sign(\sk,\allowbreak M)$ where $M$ is a message (transaction) to be signed and $\sigma_{\ECDSA}$ is a ECDSA signature. The verification algorithm is denoted as $1/0 \leftarrow \ECDSA.\Verify\allowbreak(\vk, \sigma_{\ECDSA}, M)$. 

\section{Nominative Signatures}
In this section, we give the definition of a nominative signature scheme given by Hanaoka and Schuldt~\cite{HanaokaS16} that formalizes not only the security of nominative signatures but also covers the conversion procedure that converts a nominative signature to a publicly verifiable signature. Since a smart contract cannot run the interactive verification protocols (See Section~\ref{ProposedSystem}), we need to employ the conversion functionality. Due to this reason, we employ the Hanoka-Schuldt's definition and scheme in this paper. 

\subsection{Syntax}

\begin{definition}[Syntax of Nominative Signatures~\cite{HanaokaS16}]~

\begin{description}
\item[$\Setup$:] 
The setup algorithm takes a security parameter $1^{\lambda}$ as input, and outputs a public parameter $\parameter$.

\item[$\KeyGenS$:] The signer's key generation algorithm takes $\parameter$ as input, and outputs a public/secret key pair $(\pkS, \skS)$.

\item[$\KeyGenN$:] The nominee's key generation algorithm takes $\parameter$ as input, and outputs a public/secret key pair $(\pkN, \skN)$.

\item[$\Sign$:] The signing algorithm takes $\parameter$, $\pkN$, a message to be signed $m$, and $\skS$ as input, and outputs a signature generation message $\delta$. This algorithm is run by the signer who has $\skS$. 

\item[$\Receive$:] The nominative signature generation algorithm takes $\parameter$, $\pkS$, $m$, $\delta$, and $\skN$ as input, and outputs a nominative signature $\sigma$. This algorithm is run by the nominee who has $\skN$. 

\item[$\Convert$:] The conversion algorithm takes $\parameter$, $\pkS$, $m$, $\sigma$, and $\skN$ as input, and outputs a verification token $\tk_\sigma$. This algorithm is run by the nominee who has $\skN$. 

\item[$\TkVerify$:] The token verification algorithm takes $\parameter$, $\pkS$, $\pkN$, $m$, $\sigma$, and $\tk$ as input, and outputs either $\accept$ or $\reject$. Anyone can run the algorithm because it does not take a secret key as input. 

\item[($\Comfirm, \VC$):] The interactive protocol for nominative signature confirmation takes as input $\parameter$, $\pkS$, $\pkN$, $m$, 
 and $\sigma$ as common input, the $\Comfirm$ algorithm takes $\skN$ as input, and outputs either $\accept$ or $\reject$. This protocol is run by the nominee who has $\skN$ and the verifier. 

\item[($\Disavow, \VD$):] The interactive protocol for
nominative signature disavowal takes $\parameter$, $\pkS$, $\pkN$, $m$, 
 and $\sigma$ as common input, the $\Disavow$ algorithm takes $\skN$ as input, and outputs either $\accept$ or $\reject$. This protocol is run by the nominee who has $\skN$ and the verifier. 
\end{description}
\end{definition}
\begin{figure}[t]
\centering
\includegraphics[width=9cm]{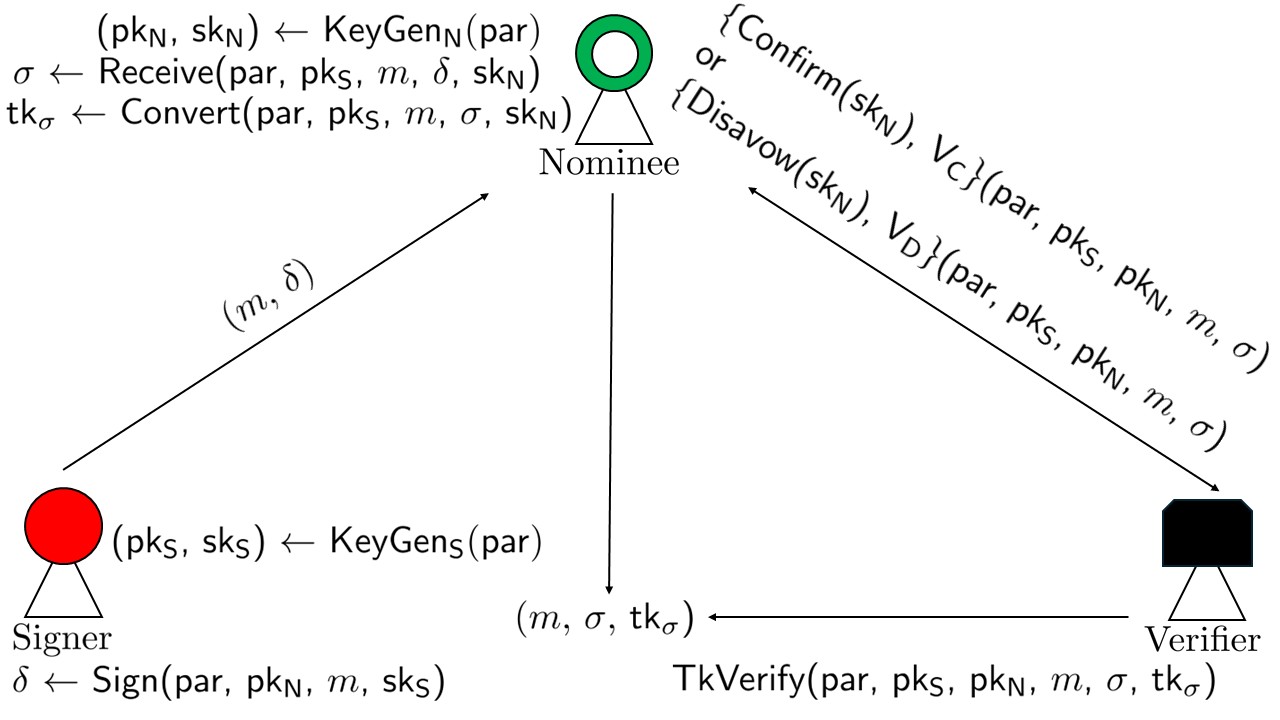}
\caption{Nominative Signatures}
\label{fig:outline}
\end{figure}

\noindent
An outline of a nominative signature scheme is as follows (See Fig.~\ref{fig:outline}).
Let $\parameter \leftarrow \Setup(1^{\lambda})$. 
A signer runs $(\pkS, \skS) \allowbreak\leftarrow \KeyGenS\allowbreak(\parameter)$. A nominee runs $(\pkN, \skN) \allowbreak\leftarrow \KeyGenN(\parameter)$. A signer and a nominee collaboratively generate a nominative signature on a message $m$ as follows. The signer generates a signature generation message $\delta \leftarrow \Sign(\parameter, \pkN, m, \skS)$, and sends $m$ and  $\delta$ to the nominee. 
The nominee generates a nominative signature $\sigma \leftarrow \Receive(\sf par, \pkS, m, \delta, \skN)$.  
To verify $(m,\sigma)$, the nominee and a verifier run $\{\Comfirm(\skN), \VC \}\allowbreak(\parameter, \allowbreak \pkS, \allowbreak\pkN, m, \sigma)$ or $\{\Disavow(\skN), \VD \}\allowbreak(\parameter, \allowbreak \pkS, \allowbreak\pkN, m, \sigma)$. 
The nominee can convert $\sigma$ to a publicly verifiable signature (called a token) $\tk_{\sigma} \leftarrow \Convert(\parameter, \pkS, m, \sigma, \skN)$. Anyone can verify $\tk_\sigma$ by running $\TkVerify(\parameter, \pkS, \pkN, m, \sigma,\allowbreak \tk_{\sigma})$. 

\subsection{Security of Nominative Signatures}

Here, we briefly introduce the security of nominative signatures (See~\cite{HanaokaS16} for more details) and briefly introduce the roles of each security in the proposed system. 

\begin{description}
    \item[Invisibility.] It guarantees that even a malicious signer cannot distinguish between an honestly generated nominative signature and a random value. Therefore, an adversary who has the signer's secret key $\skS$ and other information, and even a verifier who has once participated in the verification cannot know the correspondence between the message and the nominative signature. In the proposed system, this security is employed to guarantee that the information of business operator and investor is not leaked when nominative signatures are stored on the blockchain.

\smallskip
    \item[Unforgeability.] It guarantees that a legitimate nominative signature can be obtained only through a signer. In other words, even a malicious nominee cannot generate a nominative signature without communicating with the signer. In the proposed system, this security is employed to ensure that the investor cannot independently generate legitimate nominative signatures associated with the business operator, i.e., it guarantees that both the business operator and the investor have agreed to produce nominative signatures (together with Security against malicious signers as described below).

\smallskip
    \item[Security against malicious signers.] It guarantees that a legitimate nominative signature is produced only when the nominee is participated in. In other words, even a malicious signer cannot generate a legitimate nominative signature without communicating with the nominee. In addition, it guarantees that the signer cannot generate legitimate verification tokens and cannot prove the verification result to a third party through an interactive protocol. In the proposed system, this security is employed to ensure that the business operator cannot generate a legitimate nominative signature related to the investor by itself.

\smallskip
    \item[Protocol Security.] It also was referred as non-transferability. It guarantees that even a verifier who has executed an interactive protocol with the nominee and has verified the signature cannot prove the verification result to a third party. To satisfy the property,  the interactive protocol needs to be zero-knowledge. In the proposed system, this security guarantees that a capitalist whose investor certifies the verification result of a nominative signature cannot prove the verification result to a third party.
\end{description}

\begin{figure*}[t]
\centering
\includegraphics[width=14cm]{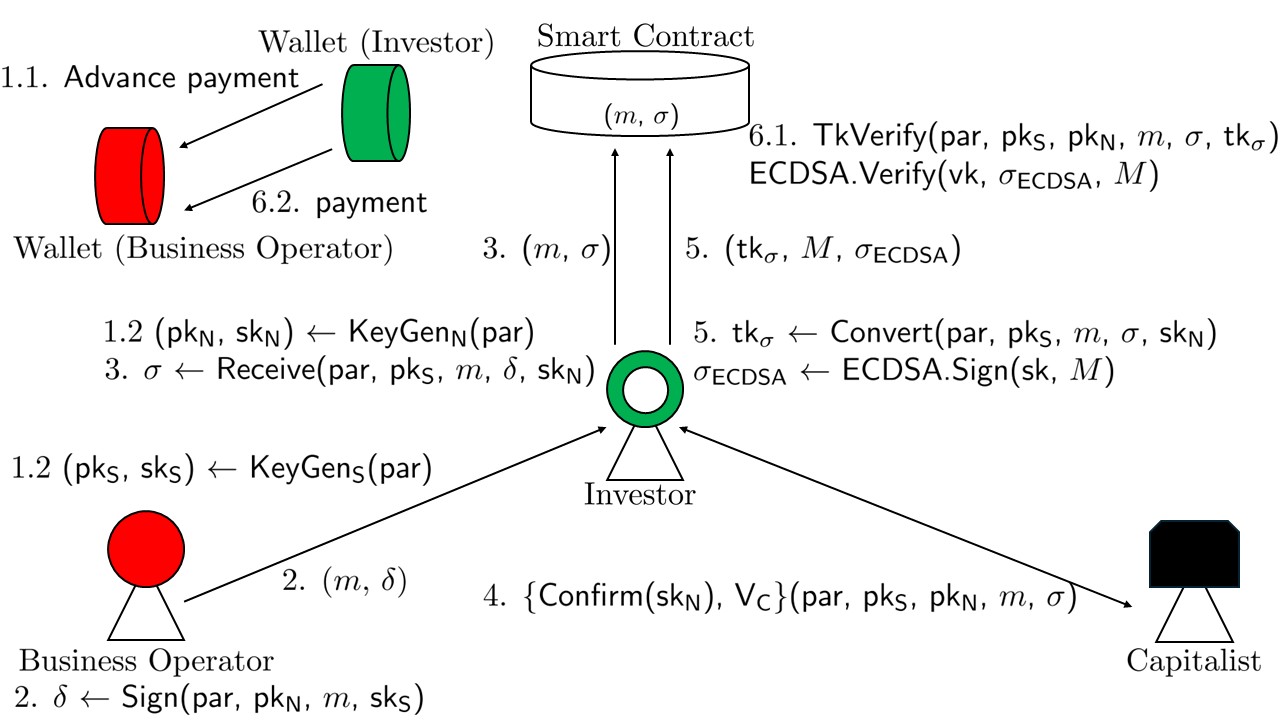}
\caption{Proposed System}
\label{fig:system}
\end{figure*}

\section{Proposed System}\label{ProposedSystem}
In this section, we propose a smart contract-based non-transferable signature verification system using nominative signatures. 
In the proposed system, the business operator acts as the signer and the investor acts as the nominee. They generate a nominative signature $\sigma$ on the program source code of the smart contract, and store $\sigma$ in a smart contract. Due to the invisibility, information of the business operator and the investor is not leaked even $\sigma$ is stored on the public blockchain. 
Note that the verifier needs to choose a random number in the interactive verification protocols of the underlying nominative signature scheme. Thus, the smart contact is not allowed to run these protocols as the verifier because the random number is disclosed when the smart contract runs these protocols. 
Thus, we employ these interactive protocols when the investor proves the validity of $\sigma$ to the capitalist off-chain. 
Tu run the smart contract, the investor converts $\sigma$ to $\tk_\sigma$ via the $\Convert$ algorithm and sends $\tk_\sigma$ to the smart contract that can be regarded as the trigger of the contract because $\tk_\sigma$ is a publicly verifiable. 
Note that a ECDSA signature is required for executing a transaction. Thus, the investor sends a ECDSA signature $\sigma_{\ECDSA}$ together with $\tk_\sigma$. 

Before giving the proposed system, we consider the cases that the investor does not follow the procedure. 
First of all, we need to consider the case that the investor does not send $(\tk_\sigma, \sigma_{\ECDSA})$ to the smart contract (or the case that $(\tk_\sigma, \sigma_{\ECDSA})$ sent to the smart contract is invalid). Then, the smart contract does not transfer the investment to the business operator. 
To capture the case, a portion of the investment amount is paid in advance in the proposed system. The investor makes an advance payment to the wallet of the business operator, which is confirmed by the business operator, who then generates the signature generation message $\delta$ and sends $\delta$ to the inverter. 
Here, we do not consider the case that $\tk_\sigma$ is invalid but $\sigma_{\ECDSA}$ is valid (then, the investment is transferred without employing the smart contact by anyone who obtains $\sigma_{\ECDSA}$) because there is no merit of the investor.

\subsection{System Description}

We give the proposed non-Transferable signature verification system using smart contracts and nominative signatures as follows (See Fig. \ref{fig:system}). We assume that the business operator and the investor manage own wallet. Moreover, we assume that they have agreed with the description of the program source code $m$ of the smart contact and the smart contract has already been deployed by the business operator.

\begin{description}
\item[1. Setup.]~
\begin{description}

\item[1.1. Advance Payment.] The investor pays a portion of the investment amount in advance to the business operator's wallet. The business operator will not process any further transactions if the advance payment is not transferred or is insufficient amount. 


\smallskip
\item[1.2. Key Generation.]. Let $\parameter \leftarrow \Setup(1^{\lambda})$. 
The business operator runs $(\pkS, \skS) \allowbreak\leftarrow \KeyGenS\allowbreak(\parameter)$, and the investor runs $(\pkN, \skN) \allowbreak\leftarrow \KeyGenN(\parameter)$.
\end{description}

\smallskip
\item[2. Signing by the business operator.]  
The business operator runs $\delta \leftarrow \Sign(\parameter, \pkN, m, \skS)$, and then sends the signature generation message $\delta$ to the investor. 

\smallskip
\item[3. Signing by the investor.]  
The investor runs $\sigma \leftarrow \Receive\allowbreak(\sf par, \pkS, m, \delta, \skN)$, and then sends $\sigma$ to the smart contract.

\smallskip
\item[4. Advertisement.] The business operator requests the investor to advertise the investment. The investor and the capitalist run $\{\Comfirm(\skN), \VC \}\allowbreak(\parameter, \allowbreak \pkS, \allowbreak\pkN, m, \sigma)$ where the investor acts as the nominee and the capitalist acts as the verifier. 

\smallskip
\item[5. Trigger Generation.] Let $M$ be the transaction. 
The investor runs $\tk_{\sigma} \leftarrow \Convert(\parameter, \pkS, m, \sigma, \skN)$ and 
$\sigma_{\sf ECDSA} \allowbreak\leftarrow \ECDSA.\Sign (\sk, M)$, and sends $\tk_{\sigma}$ and $(M,\sigma_{\ECDSA})$ to the management smart contract. 

\smallskip
\item[6. Investment]~
\begin{description}

\item[6.1. Trigger Verification]
$\TkVerify(\parameter, \pkS, \pkN, m,\allowbreak \sigma,\allowbreak \tk_{\sigma})$ and $\ECDSA.\Verify\allowbreak(\vk, \sigma_{\ECDSA},\allowbreak M)$ are run in the management smart contract.

\smallskip
\item[6.2. Execution]
If $(\tk_{\sigma},\sigma_{\ECDSA})$ is valid, the funds are transferred from the investor's wallet to the business operator's wallet according to the program $m$ and transaction $M$ 
If either $\tk_{\sigma}$ or $\sigma_{\ECDSA}$ is not valid, the smart contract does not transfer funds. 
\end{description}

\end{description}

\subsection{Security of Proposed System}

Due to the unforgeability of the underlying nominative signature scheme, the proposed system guarantees that investors cannot generate legitimate nominative signatures without communicating with the business operator. Moreover, due to the security against malicious signers, the business operator cannot generate legitimate nominative signatures without communicating with the investor. 
Therefore, it is guaranteed that it is impossible to forge a legitimate nominative signature without the agreement of both parties. 
Due to the invisibility, the information of the business operator and the investor cannot be leaked from the nominative signature $\sigma$ stored on the public blockchain. Moreover, due to the non-transferability, the capitalist who has known the verification result cannot prove the result to a third party.

\section{Nominative Signatures over Asymmetric Pairings}
\label{NSoverAP}

In this section, we transform the Hanaoka-Schuldt nominative signature scheme to employ asymmetric pairings. For the sake of clarity, we introduced the original scheme as follows. Hereafter, we denote the zero-knowledge proof of the relation $R$ between the witness $\omega$ and the common input $x$ as $\ZKPK\{(\omega):R(x,\omega)\}$. With respect to the $\Comfirm/\Disavow$ protocol, we use a four-pass interactive zero-knowledge proof constructed using the Cramer-Damg{\aa}rd-MacKenzie transformation~\cite{CramerDM00} (Nakagawa et al.~\cite{NakagawaNOEHSK17} described the actual procedure of the protocols). 
In the evaluation in Section~\ref{PerformanceEvaluation}, we set $\ell=256$ for the plaintext space $\{0,1\}^\ell$. In addition, we assumed that $F_S(M_S)$ and $F_N(M_N)$ are run by 128-times additions over elliptic curves on average, respectively, and that the estimation of the total number of additions is 256. 

\medskip
\noindent\textbf{Original Hanaoka-Schuldt Nominative Signature Scheme}
\begin{description}
\item[$\Setup(1^{\lambda})$:]
Let $\G_1$ and $\G_T$ be groups with prime order $p$, $g$ be the generator of $\G_1$, and $e:\G_1 \times \G_1 \rightarrow \G_T$ be a symmetric pairing. 
Choose collision-resistant hash functions $H_1 :\{0,1\}^* \rightarrow \{0,1\}^\ell$ and $H_2 : \{0,1\}^* \rightarrow \mathbb{Z}_p$. Output $\parameter \leftarrow (e,p,g,H_1,H_2)$. Here, $\{0,1\}^\ell$ is the plaintext space. 

\smallskip
\item[$\KeyGenS(\parameter)$:]
Choose $\alpha_S, v_0, \ldots, v_\ell \leftarrow \mathbb{Z}_p$, $h_S \leftarrow \mathbb{G}_1$, and compute $g_S \leftarrow g^{\alpha_S}$, $u_i \leftarrow g^{v_i}\ (0 \leq i \leq \ell)$. Let $m_i$ be the $i$-th bit of $m \in \{0,1\}^\ell$, and define the function $\FS(m) = u_0 \prod^{\ell}_{i=1} u_i^{m_i}$. Output $\pkS \leftarrow (g_S, h_S, u_0, \ldots, u_\ell)$ and $\skS \leftarrow \alpha_S$. 

\smallskip
\item[$\KeyGenN(\parameter)$:]
Choose $\alpha_N, y_1, y_2, v_0', \ldots, v_\ell' \leftarrow \mathbb{Z}_p$,
$h_N, k \leftarrow \mathbb{G}_1$, and compute $g_N \leftarrow g^{\alpha_N}$ and $u^\prime_i \leftarrow g^{v^\prime_i}\ (1 \leq i \leq \ell)$. 
Let $m_i$ be the $i$-th bit of $m \in \{0,1\}^\ell$, 
and define the function $\FN(m) = u'_0 \prod^{\ell}_{i=1} {u'}_i^{m_i}$. 
Then, compute $x_1 \leftarrow g^{y_1^{-1}}$ and $x_2 \leftarrow g^{y_2^{-1}}$. Output $\pkN \leftarrow (g_N, h_N, k, u'_0, \ldots, u'_n, x_1, x_2)$ and $\skN \leftarrow (\alpha_N, v'_0, \ldots, v'_\ell, y_1, y_2)$. 

\smallskip
\item[$\Sign(\parameter, \pkN, m, \skS)$:]
Choose $r \leftarrow \mathbb{Z}_p$, and compute $\MS=H_1(\pkN||m)$, $\delta_1 \leftarrow g^r,\delta_2 \leftarrow h_S^{\alpha_S}\FS (\MS)^r$. Then, output $\delta=(\delta_1,\delta_2)$.

\smallskip
\item[$\Receive(\parameter, \pkS, m, \delta, \skN)$:]
Compute $\MS=H_1(\pkN||m)$ and output $\bot$ if $e(g_S, h_S)e(\delta_1, \FS(\MS))=e(g, \delta_2)$ does not hold. 
Otherwise, choose $r, r', s \leftarrow \mathbb{Z}_p$ and compute $\delta_1' \leftarrow \delta_1g^{r'}, \delta_2' \leftarrow \delta_2\FS(\MS)^{r'}$. Then, compute $t \leftarrow H_2(\pkN||\sigma_1|||\sigma_2||m), M_N \leftarrow g^tk^s$. Let $M_{N,i}$ be the $i$-th bit of $\MN$. Compute $\sigma_1 \leftarrow (\delta_1'/g^r)^{y_1^{-1}}$, $\sigma_2 \leftarrow (g^r)^{y_2^{-1}}$, $\sigma_3 \leftarrow \delta_2'h_N^{\alpha_N}(\delta_1')^{v'_0+\prod^{\ell}_{i=1} v'_iM_{N,i}}$. Output $\sigma \leftarrow (\sigma_1,\sigma_2,\allowbreak \sigma_3, s)$. 

\smallskip
\item[$\Convert(\parameter, \pkS, m, \sigma, \skN)$:]
For $M_S=H_1(\pkN||m)$, $M_N=g^tk^s$, $t = H_2(\pkS||\sigma_1||\sigma_2|||m)$, output $\bot$ if
\begin{gather*}
\begin{split}
e(g, \sigma_3)=&e(g_S, h_S) \cdot e(g_N, h_N)\\
\cdot &e(\sigma_1^{y_1}\sigma_2^{y_2}, F_S(M_S)F_N(M_N))
\end{split}
\end{gather*}
does not hold. Otherwise, output $\tk_{\sigma} \leftarrow (\sigma_1^{y_1}, \sigma_2^{y_2})$.

\smallskip
\item[$\TkVerify(\parameter, \pkS, \pkN, m, \sigma, \tk_{\sigma})$:]
For $M_S=H_1(\pkN||m)$, $M_N=g^tk^s$, $\tk_{\sigma}=(\tk_1,\tk_2)$, 
$t = H_2(\pkS||\sigma_1||\sigma_2||m)$,  output $\accept$ if
\begin{gather*}
\begin{split}
e(\sigma _1, g)=&e(\tk_1, x_1)\\ 
e(\sigma _2, g)=&e(\tk_2, x_2)\\
e(g, \sigma_3)=&e(g_S, h_S)\cdot e(g_N, h_N) \\
\cdot & e(\tk_1 \tk_2, \FS (\MS) \FN (\MN) ) 
\end{split}
\end{gather*}
holds and $\reject$, otherwise

\smallskip
\item[$(\Comfirm(\skN), V_C)(\parameter, \pkS, \pkN, m, \sigma)$:]
For $M_S=H_1\allowbreak(\pkN||m)$, $M_N=g^tk^s$, $t = H_2(\pkS||\sigma_1||\sigma_2||m)$, define $e_1=e(g,\sigma_3)$, $e_2=e(g_S, h_S) e(g_N, h_N)$, $e_3=e(\sigma_1,\allowbreak F_S(M_S)F_N(M_N))$, $e_4=e(\sigma_2, F_S(M_S)F_N(M_N))$. The following protocol is executed between the nominee and a verifier. 
$$\ZKPK\{(y_1, y_2):x_1^{y_1}=g \land x_2^{y_2}=g \land e_1=e_2e_3^{y_1}e_4^{y_2} \}$$

\item[$(\Disavow(\skN), V_D)(\parameter, \pkS, \pkN, m, \sigma)$:]
The following protocol is executed between the nominee and a verifier. 
$$\ZKPK\{(y_1, y_2):x_1^{y_1} = g \land x_2^{y_2} = g \land e_1 \neq e_2e_3^{y_1}e_4^{y_2} \}$$

\end{description}

\noindent
\textbf{What we need to consider for transformation?} Let $\G_1\times\G_2\rightarrow\G_T$ be an asymmetric pairing where $\G_1$, $\G_2$, and $\G_T$ are groups with prime order $p$ and $g_1\in\G_1$ and $g_2\in\G_2$ are generators. We consider type 3 curves where no efficient isomorphism between $\G_1$ and $\G_2$ exist. 

If each element belongs to $\G_1$ or $\G_2$, then the transformation is easy. However, an element belongs to both groups in the Hanaoka-Schuldt nominative signature scheme if we naively assign each element. Concretely, for $$\sigma_3 \leftarrow \delta_2'h_N^{\alpha_N}(\delta_1')^{v'_0+\prod^{\ell}_{i=1} v'_iM_{N,i}}$$ 
$\sigma_3$ must belong to $\G_2$ because $e(g, \sigma_3)$ is computed. However, for $\delta=(\delta_1,\delta_2)=(g^r,h_S^{\alpha_S}\FS (\MS)^r)$ (which is a Waters signature~\cite{Waters05} on $\MS$), $\delta_1$ must belong to $\G_1$ and $\delta_2$ must belong to $\G_2$, respectively, because of the verification equation $e(g_S, h_S)e(\delta_1, \FS(\MS))=e(g, \delta_2)$. Since $\delta_1' \leftarrow \delta_1g^{r'}$,  $\delta_1'\in \G_1$. This contradicts the computation above. If we assign $\sigma_3$ to $\G_1$ and refine  $e(g, \sigma_3)$ to be $e( \sigma_3,g)$, then $\delta_2$ must belong to $\G_1$ since $\delta_2' \leftarrow \delta_2\FS(\MS)^{r'}$. Then, $\FS(\MS)\in\G_1$ and thus $h_S\in\G_1$ because $\delta_2=h_S^{\alpha_S}\FS (\MS)^r$. This contradicts the computation $e(g_S, h_S)$. 

\medskip
\noindent
\textbf{Our Modification}. We duplicate $\delta_1$ such that $\delta=(\delta_1,\delta_2,\delta_3)=(g_1^r,g_2^r,h_S^{\alpha_S}\FS (\MS)^r)\in\G_1\times\G_2^2$ (here, previous $\delta_2$ is renamed as $\delta_3$). Then, $$\sigma_3 \leftarrow \delta_3'h_N^{\alpha_N}(\delta_2')^{v'_0+\prod^{\ell}_{i=1} v'_iM_{N,i}}\in\G_2$$ 
where $\delta_2' \leftarrow \delta_2g_2^{r'}\in\G_2$ and $\delta_3' \leftarrow \delta_3\FS(\MS)^{r'}\in\G_2$. 
Moreover, we can run the verification equation: $$e(g_S, h_S)e(\delta_1, \FS(\MS))=e(g_1, \delta_3)$$
\noindent 
To guarantee the security of the modified scheme, three additional problems happen. First, we additionally need to check $$\log_{g_1}\delta_1=\log_{g_2}\delta_2$$ because the same randomness is used for computing $\delta_1$ and $\delta_2$. 
This can be checked via the pairing operation: $$e(\delta_1,g_2)=e(g_1,\delta_2)$$ holds or not. This is the reason behind that we will add the equation in the $\Receive$ algorithm later. 
The second problem is more serious. In the security proof of the original scheme, the security is reduced to the Waters signature scheme. In the symmetric pairing setting, the signing oracle of the Waters signature scheme returns $(g^r,h_S^{\alpha_S}\FS (\MS)^r)$ for the signing query $\MS$. If we simply construct the Waters signature scheme over $\G_2$, then the signing oracle returns $(g_2^r,h_S^{\alpha_S}\FS (\MS)^r)\in\G_2^2$. However, the simulator has no way to produce $g_1^r$ without knowing $r$. Thus, the simulator needs to receive $(g_1^r,g_2^r,h_S^{\alpha_S}\FS (\MS)^r)\in\G_1\times\G_2^2$ from the signing oracle. That is, we need to modify the Waters signature scheme. 
The third problem is how to modify the decision linear (DLIN) assumption that is also employed to prove the security of the original scheme. 

We describe the modified nominative signature scheme first as follows, and we evaluate the modified Waters signature scheme and the DLIN problem later. 
Compared to the original scheme, our modification does not increase the number of pairing computations in the $\TkVerify$ algorithm that is run on the smart contract. 

\medskip
\noindent\textbf{Modified Hanaoka-Schuldt Nominative Signature Scheme over Asymmetric Pairings}
\begin{description}
\item[$\Setup(1^{\lambda})$:]
Let $\G_1\times\G_2\rightarrow\G_T$ be an asymmetric pairing where $\G_1$, $\G_2$, and $\G_T$ are groups with prime order $p$ and $g_1\in\G_1$ and $g_2\in\G_2$ are generators. 
Choose collision-resistant hash functions $H_1 :\{0,1\}^* \rightarrow \{0,1\}^\ell$ and $H_2 : \{0,1\}^* \rightarrow \mathbb{Z}_p$. Output $\parameter \leftarrow (e,p,g_1,g_2,H_1,H_2)$. Here, $\{0,1\}^\ell$ is the plaintext space. 

\smallskip
\item[$\KeyGenS(\parameter)$:]
Choose $\alpha_S, v_0, \ldots, v_\ell \leftarrow \mathbb{Z}_p$, $h_S \leftarrow \mathbb{G}_2$, and compute $g_S \leftarrow g_1^{\alpha_S}\in\G_1$, $u_i \leftarrow g_2^{v_i}\in\G_2\ (0 \leq i \leq \ell)$. Let $m_i$ be the $i$-th bit of $m \in \{0,1\}^\ell$, and define the function $\FS(m) = u_0 \prod^{\ell}_{i=1} u_i^{m_i}\in\G_2$. Output $\pkS \leftarrow (g_S, h_S, u_0, \ldots, u_\ell)$ and $\skS \leftarrow \alpha_S$. 

\smallskip
\item[$\KeyGenN(\parameter)$:]
Choose $\alpha_N, y_1, y_2, v_0', \ldots, v_\ell' \leftarrow \mathbb{Z}_p$,
$h_N, k \leftarrow \mathbb{G}_2$, and compute $g_N \leftarrow g_1^{\alpha_N}\in\G_1$ and $u^\prime_i \leftarrow g_2^{v^\prime_i}\in\G_2\ (1 \leq i \leq \ell)$. 
Let $m_i$ be the $i$-th bit of $m \in \{0,1\}^\ell$, 
and define the function $\FN(m) = u'_0 \prod^{\ell}_{i=1} {u'}_i^{m_i}\in\G_2$. 
Then, compute $x_1 \leftarrow g_2^{y_1^{-1}}\in\G_2$ and $x_2 \leftarrow g_2^{y_2^{-1}}\in\G_2$. Output $\pkN \leftarrow (g_N, h_N, k, u'_0, \ldots, u'_n, x_1, x_2)$ and $\skN \leftarrow (\alpha_N, v'_0, \ldots, v'_\ell, y_1, y_2)$. 

\smallskip
\item[$\Sign(\parameter, \pkN, m, \skS)$:]
Choose $r \leftarrow \mathbb{Z}_p$, and compute $\MS=H_1(\pkN||m)$, $\delta_1 \leftarrow g_1^r\in\G_1$, $\delta_2\leftarrow g_2^r\in\G_2$, and 
$\delta_3 \leftarrow h_S^{\alpha_S}\FS (\MS)^r\in\G_2$. Then, output $\delta=(\delta_1,\delta_2,\delta_3)$.

\smallskip
\item[$\Receive(\parameter, \pkS, m, \delta, \skN)$:]
Compute $\MS=H_1(\pkN||m)$ and output $\bot$ if both $e(g_S, h_S)e(\delta_1, \FS(\MS))=e(g_1, \delta_2)$ and $e(\delta_1,g_2)=e(g_1,\delta_2)$ do not hold. 
Otherwise, choose $r, r', s \leftarrow \mathbb{Z}_p$ and compute $\delta_1' \leftarrow \delta_1g_1^{r'}\in\G_1$, $\delta_2' \leftarrow \delta_2g_2^{r'}\in\G_2$, and 
$\delta_3' \leftarrow \delta_3\FS(\MS)^{r'}$. Then, compute $t \leftarrow H_2(\pkN||\sigma_1|||\sigma_2||\sigma_3||m), M_N \leftarrow g_2^tk^s$. Let $M_{N,i}$ be the $i$-th bit of $\MN$. Compute $\sigma_1 \leftarrow (\delta_1'/g_1^r)^{y_1^{-1}}\in\G_1$, $\sigma_2 \leftarrow (g_1^r)^{y_2^{-1}}\in\G_1$, and $\sigma_3 \leftarrow \delta_3'h_N^{\alpha_N}(\delta_2')^{v'_0+\prod^{\ell}_{i=1} v'_iM_{N,i}}\in\G_2$. Output $\sigma \leftarrow (\sigma_1,\sigma_2,\allowbreak \sigma_3, s)$. 

\smallskip
\item[$\Convert(\parameter, \pkS, m, \sigma, \skN)$:]
For $M_S=H_1(\pkN||m)$, $M_N=g_2^tk^s$, $t = H_2(\pkS||\sigma_1||\sigma_2||\sigma_3|||m)$, output $\bot$ if  
\begin{gather*}
\begin{split}
e(g_1, \sigma_3)=&e(g_S, h_S) \cdot e(g_N, h_N)\\
\cdot &e(\sigma_1^{y_1}\sigma_2^{y_2}, F_S(M_S)F_N(M_N))
\end{split}
\end{gather*}
does not hold. Otherwise, output $\tk_{\sigma} \leftarrow (\sigma_1^{y_1}, \sigma_2^{y_2})\in\G_1^2$.

\smallskip
\item[$\TkVerify(\parameter, \pkS, \pkN, m, \sigma, \tk_{\sigma})$:]
For $M_S=H_1(\pkN||m)$, $M_N=g_2^tk^s$, $\tk_{\sigma}=(\tk_1,\tk_2)$, 
$t = H_2(\pkS||\sigma_1||\sigma_2||\sigma_3||m)$,  output $\accept$ if
\begin{gather*}
\begin{split}
e(\sigma_1, g_2)=&e(\tk_1, x_1)\\ 
e(\sigma_2, g_2)=&e(\tk_2, x_2)\\
e(g_1, \sigma_3)=&e(g_S, h_S)\cdot e(g_N, h_N) \\
\cdot & e(\tk_1 \tk_2, \FS (\MS) \FN (\MN) ) 
\end{split}
\end{gather*}
holds and $\reject$, otherwise

\smallskip
\item[$(\Comfirm(\skN), V_C)(\parameter, \pkS, \pkN, m, \sigma)$:]
For $M_S=H_1\allowbreak(\pkN||m)$, $M_N=g_2^tk^s$, $t = H_2(\pkS||\sigma_1||\sigma_2||\sigma_3||m)$, define $e_1=e(g_1,\sigma_3)$, $e_2=e(g_S, h_S) e(g_N, h_N)$, $e_3=e(\sigma_1,\allowbreak F_S(M_S)F_N(M_N))$, $e_4=e(\sigma_2, F_S(M_S)F_N(M_N))$. The following protocol is executed between the nominee and a verifier. 
$$\ZKPK\{(y_1, y_2):x_1^{y_1}=g_2 \land x_2^{y_2}=g_2 \land e_1=e_2e_3^{y_1}e_4^{y_2} \}$$

\item[$(\Disavow(\skN), V_D)(\parameter, \pkS, \pkN, m, \sigma)$:]
The following protocol is executed between the nominee and a verifier. 
$$\ZKPK\{(y_1, y_2):x_1^{y_1} = g_2 \land x_2^{y_2} = g_2 \land e_1 \neq e_2e_3^{y_1}e_4^{y_2} \}$$

\end{description}

\noindent
\textbf{Evaluation of the Modified Waters Signature Scheme}. What we need to evaluate here is whether the Waters signature scheme provides existentially unforgeability against chosen message attack (EUF-CMA) when a signature is modified as $$(g_1^r,g_2^r,h^{\alpha}H_W (m)^r)\in\G_1\times\G_2^2$$ where $h\in\G_2$, $\alpha\in\mathbb{Z}_p$, and $H_W:\{0,1\}\rightarrow \G_2$ is the Waters hash. 

We revisited the original security proof~\cite{Waters05}. Let $(g,g^a,g^b)\in\G^3$ be an Computational Diffie-Hellman (CDH) instance where $e: \G\times\G\rightarrow\G_T$ is a symmetric pairing. 
In the original security proof, for a signature $(g^{\tilde{r}},h^{\alpha}H_W (m)^{\tilde{r}})$, the randomness $\tilde{r}$ is implicitly set as $\tilde{r}:=r-\frac{a}{F(v)}$ to cancel out $h^\alpha=g^{ab}$ (which is the solution of the CDH problem) for computing $h^{\alpha}H_W (m)^{\tilde{r}}$. Here, $a$ is contained in the CDH instance, $\alpha:=a$, and $r$ is a randomness chosen by the simulator. 
We omit the definition of $F(v)$ since it is not necessary in the evaluation below. 

To simulate this procedure when $(g_1^{\tilde{r}},g_2^{\tilde{r}},h^{\alpha}H_W (m)^{\tilde{r}})$ is computed, we need to define a CDH problem over asymmetric bilinear groups: for $(g_1,g_1^a,g_2,g_2^a,g_2^b)\in\G_1^2\times\G_2^3$, compute $g_2^{ab}$. The modified Waters signature scheme is EUF-CMA secure under the modified CDH assumption. 

\medskip
\noindent
\textbf{Modification of the DLIN problem}. 
The original security proof employs the DLIN problem: for $(x_1,x_2,x_1^{a},x_2^{b},g^c)$, decide $c=a+b$ or not. $x_1$ and $x_2$ are directly used as is, and $\sigma_1:=x_1^{a}$ and $\sigma_2:=x_2^{b}$ (i.e., implicitly set $a:=y_1$ and $b:=y_2$). Moreover, $\sigma_3:=h_S^{\alpha_S}(g^c)^{v_0+\prod^{\ell}_{i=1} v_iM_{S,i}}h_N^{\alpha_N}(g^c)^{v'_0+\prod^{\ell}_{i=1} v'_iM_{N,i}}$ and claimed that $(\sigma_1,\sigma_2,\sigma_3,s)$ is a valid signature when $c=a+b$ (here, $s$ is picked by the simulator). To embed the instance to the modified scheme, we need to define a DLIN problem over asymmetric bilinear groups: for $(x_{1,1},x_{2,1},x_{1,1}^a,x_{2,1}^b,x_{1,2},x_{2,2},x_{1,2}^a,x_{2,2}^b,g_2^c)\in\G_1^4\times\G_2^5$, decide $c=a+b$ or not, since $\sigma_1,\sigma_2\in\G_1$ and $x_1,x_2\in\G_2$ in the modified scheme. In the simulation, set $x_1:=x_{1,2}\in\G_2$, $x_2:=x_{2,2}\in\G_2$, 
$\sigma_1:=x_{1,1}^{a}\in\G_1$,  $\sigma_2:=x_{2,1}^{b}\in\G_1$, and $\sigma_3:=h_S^{\alpha_S}(g_2^c)^{v_0+\prod^{\ell}_{i=1} v_iM_{S,i}}h_N^{\alpha_N}(g_2^c)^{v'_0+\prod^{\ell}_{i=1} v'_iM_{N,i}}\in\G_2$. 

\section{Performance Evaluation}
\label{PerformanceEvaluation}

\noindent\textbf{Gas Cost}. 
In this section, we estimate the gas cost of the proposed system when the $\textsf{TkVerify}$ algorithm is run by the smart contract. 
First, we need to select the underlying nominative signature scheme. 
Due to the progress of quantum computers, one may think that we should employ a lattice-based nominative signature scheme, e.g.,~\cite{KansalDM21}. However, to the best of our knowledge, no pre-compiled contract providing a lattice-based cryptographic scheme has been published so far. Moreover, currently, ECDSA is necessary to issue a transaction which is secure under the discrete-logarithm problem over elliptic curves~\cite{FerschKP16,GrothS22} and is not a post-quantum cryptography (PQC). Thus, replacing the underlying nominative signature scheme to be PQC does not affect the post-quantum security of the proposed system. Thus, we employ the pairing-based Hanaoka-Schuldt nominative signature scheme~\cite{HanaokaS16} in our evaluation. 

We employed a pre-compiled contract provided by smlXL.Inc~\cite{smlXL} that allows us to run the smart contract efficiently. 
It provides elliptic curve operations and pairing computation on Barreto-Naehrig (BN) curves~\cite{BarretoN05} (bn128). 
First, we introduce a benchmark of pairing computations and additions over elliptic curves in Table \ref{smlXL}.  

\begin{table}[ht]
\centering
\caption{Process and gas costs}
\label{smlXL}
\begin{tabular}{|l|l|} \hline
Process & Gas cost (Unit) \\ \hline
Pairing & $45,000 + 34,000 * n$ \\ 
Addition over elliptic curves & 150 \\ \hline
\end{tabular}
\end{table}

\noindent 
Here,  $n$ is the number of pairing computations. Precisely, $n$ increases for every 192 bytes of input size. In the Hanaoka-Schuldt nominative signature scheme, eight pairing computations are required in the $\textsf{TkVerify}$ algorithm. 
By considering the gas cost in table \ref{smlXL}, the gas cost for running the $\textsf{TkVerify}$ algorithm is estimated to be 317,000 + 150 * 256 = 355,400 Units (0.00629058 ETH (11.2 U.S. dollars) by the rate on March 14, 2025)). 
The gas cost for verifying a ECDSA signature on the pre-compiled contract is just 3,000 Units (when the algorithm that recovers the verification key from the ECDSA signature and the message, called ${\sf ecRecover}$, is executed). 
It should be noted that running the ${\sf ecRecover}$ algorithm is an essential procedure in Ethereum, and it can be assumed that there is a consensus that this level of gas cost is acceptable to run smart contracts. 

Our estimation indicates that the execution of the $\textsf{TkVerify}$ algorithm requires about 120 times higher gas cost than that of the key recovery process of ECDSA, and we cannot say that the proposed system is efficient in practice. Nevertheless, we claim that our proposal is meaningful to demonstrate the feasibility of a smart contract-based system when enhanced cryptographic primitives such as nominative signatures are employed. 

\section{Conclusion}

In this paper, we proposed a non-transferable signature verification system. We employed both the smart contract and a nominative signature scheme, and estimate the gas cost for running the system. Currently, the proposed system is not sufficiently practical due to the number of pairing computations of the underlying nominative signature scheme. 
Proposing an efficient nominative signature scheme is also an important future work since eight pairing computations are dominant of the gas cost. In addition, it is also a future work to enable lattice-based nominative signatures to be handled on smart contracts that may reduce the gas cost.

\medskip
\noindent\textbf{Acknowledgment}: 
This work was supported by JSPS KAKENHI Grant Number JP25H01106. 


\begin{thebibliography}{10}

\bibitem{KPW96}
S.~Kim, S.~Park, and D.~Won, ``Zero-knowledge nominative signatures,''
  PragoCrypt, pp.380--392, 1996.

\bibitem{smlXL}
smlXL.Inc, ``Evm codes - precompiled contracts.''
\newblock \url{https://www.evm.codes/precompiled?fork=shanghai}.

\bibitem{HanaokaS16}
G.~Hanaoka and J.C.N. Schuldt, ``Convertible nominative signatures from
  standard assumptions without random oracles,'' {IEICE} Transactions on
  Fundamentals of Electronics, Communications and Computer Sciences, vol.99-A,
  no.6, pp.1107--1121, 2016.

\bibitem{NishinoOE25}
H.~Nishino, K.~Omote, and K.~Emura, ``A smart contract-based non-transferable
  signature verification system using nominative signatures,'' {AsiaJCIS},
  2025, to appear.

\bibitem{BarretoN05}
P.S.L.M. Barreto and M.~Naehrig, ``Pairing-friendly elliptic curves of prime
  order,'' Selected Areas in Cryptography, pp.319--331, 2005.

\bibitem{GalbraithPS08}
S.D. Galbraith, K.G. Paterson, and N.P. Smart, ``Pairings for cryptographers,''
  Discret. Appl. Math., vol.156, no.16, pp.3113--3121, 2008.

\bibitem{OzdemirB22}
A.~Ozdemir and D.~Boneh, ``Experimenting with collaborative zk-{SNARK}s:
  Zero-knowledge proofs for distributed secrets,'' {USENIX} Security Symposium,
  ed.~K.R.B. Butler and K.~Thomas, pp.4291--4308, 2022.

\bibitem{ParnoHG016}
B.~Parno, J.~Howell, C.~Gentry, and M.~Raykova, ``Pinocchio: nearly practical
  verifiable computation,'' Communications of the {ACM}, vol.59, no.2,
  pp.103--112, 2016.

\bibitem{BunzBBPWM18}
B.~B{\"{u}}nz, J.~Bootle, D.~Boneh, A.~Poelstra, P.~Wuille, and G.~Maxwell,
  ``Bulletproofs: Short proofs for confidential transactions and more,'' {IEEE}
  Symposium on Security and Privacy, pp.315--334, 2018.

\bibitem{LiuWW04}
J.K. Liu, V.K. Wei, and D.S. Wong, ``Linkable spontaneous anonymous group
  signature for ad hoc groups (extended abstract),'' {ACISP}, pp.325--335,
  2004.

\bibitem{LiuW05}
J.K. Liu and D.S. Wong, ``Linkable ring signatures: Security models and new
  schemes,'' Computational Science and Its Applications, pp.614--623, 2005.

\bibitem{ChinEO25}
K.~Chin, K.~Emura, and K.~Omote, ``An anonymous yet accountable contract wallet
  system using account abstraction,'' Journal of Information Security and
  Applications, vol.89, p.103978, 2025.

\bibitem{BootleCCGGP15}
J.~Bootle, A.~Cerulli, P.~Chaidos, E.~Ghadafi, J.~Groth, and C.~Petit, ``Short
  accountable ring signatures based on {DDH},'' {ESORICS}, ed.~G.~Pernul,
  P.Y.A. Ryan, and E.R. Weippl, pp.243--265, 2015.

\bibitem{BonehGLS03}
D.~Boneh, C.~Gentry, B.~Lynn, and H.~Shacham, ``Aggregate and verifiably
  encrypted signatures from bilinear maps,'' {EUROCRYPT}, pp.416--432, 2003.

\bibitem{P17}
A.~Poelstra, ``Scriptless scripts.'' \url{https://tinyurl.com/ludcxyz}, 2017.

\bibitem{GerhartSST24}
P.~Gerhart, D.~Schr{\"{o}}der, P.~Soni, and S.A.K. Thyagarajan, ``Foundations
  of adaptor signatures,'' {EUROCRYPT}, pp.161--189, 2024.

\bibitem{DaiOY22}
W.~Dai, T.~Okamoto, and G.~Yamamoto, ``Stronger security and generic
  constructions for adaptor signatures,'' {INDOCRYPT}, pp.52--77, 2022.

\bibitem{VanjaniST24}
N.~Vanjani, P.~Soni, and S.A.K. Thyagarajan, ``Functional adaptor signatures:
  Beyond all-or-nothing blockchain-based payments,'' {ACM} {CCS},
  pp.1493--1507, 2024.

\bibitem{0001BBKM19}
A.~Miller, I.~Bentov, S.~Bakshi, R.~Kumaresan, and P.~McCorry, ``Sprites and
  state channels: Payment networks that go faster than lightning,'' Financial
  Cryptography and Data Security, pp.508--526, 2019.

\bibitem{MadathilTVFMM23}
V.~Madathil, S.A.K. Thyagarajan, D.~Vasilopoulos, L.~Fournier, G.~Malavolta,
  and P.~Moreno{-}Sanchez, ``Cryptographic oracle-based conditional payments,''
  {NDSS}, The Internet Society, 2023.

\bibitem{QinPMSESELYY23}
X.~Qin, S.~Pan, A.~Mirzaei, Z.~Sui, O.~Ersoy, A.~Sakzad, M.F. Esgin, J.K. Liu,
  J.~Yu, and T.H. Yuen, ``{BlindHub}: Bitcoin-compatible privacy-preserving
  payment channel hubs supporting variable amounts,'' {IEEE} Symposium on
  Security and Privacy, pp.2462--2480, 2023.

\bibitem{ChaumA89}
D.~Chaum and H.V. Antwerpen, ``Undeniable signatures,'' CRYPTO, pp.212--216,
  1989.

\bibitem{BoyarCDP90}
J.~Boyar, D.~Chaum, I.~Damg{\aa}rd, and T.P. Pedersen, ``Convertible undeniable
  signatures,'' CRYPTO, pp.189--205, 1990.

\bibitem{Chaum94}
D.~Chaum, ``Designated confirmer signatures,'' EUROCRYPT, pp.86--91, 1994.

\bibitem{LiskovM08}
M.~Liskov and S.~Micali, ``Online-untransferable signatures,'' Public Key
  Cryptography, pp.248--267, 2008.

\bibitem{JakobssonSI96}
M.~Jakobsson, K.~Sako, and R.~Impagliazzo, ``Designated verifier proofs and
  their applications,'' EUROCRYPT, pp.143--154, 1996.

\bibitem{HuangLW08}
Q.~Huang, D.Y.W. Liu, and D.S. Wong, ``An efficient one-move nominative
  signature scheme,'' International Journal of Applied Cryptography, vol.1,
  no.2, pp.133--143, 2008.

\bibitem{LiuWHWHMS07}
D.Y.W. Liu, D.S. Wong, X.~Huang, G.~Wang, Q.~Huang, Y.~Mu, and W.~Susilo,
  ``Formal definition and construction of nominative signature,'' {ICICS},
  pp.57--68, 2007.

\bibitem{ZhaoY09}
W.~Zhao and D.~Ye, ``Pairing-based nominative signatures with selective and
  universal convertibility,'' Inscrypt, pp.60--74, 2009.

\bibitem{HuangW04}
Z.~Huang and Y.~Wang, ``Convertible nominative signatures,'' {ACISP},
  pp.348--357, 2004.

\bibitem{LiuCWM07}
D.Y.W. Liu, S.~Chang, D.S. Wong, and Y.~Mu, ``Nominative signature from ring
  signature,'' {IWSEC}, pp.396--411, 2007.

\bibitem{SchuldtH11}
J.C.N. Schuldt and G.~Hanaoka, ``Non-transferable user certification secure
  against authority information leaks and impersonation attacks,'' ACNS,
  pp.413--430, 2011.

\bibitem{NakagawaEHKNOS14}
S.~Nakagawa, K.~Emura, G.~Hanaoka, A.~Kodate, T.~Nishide, E.~Okamoto, and
  Y.~Sakai, ``A privacy-enhanced access log management mechanism in {SSO}
  systems from nominative signatures,'' {IEEE} {TrustCom}, pp.565--574, 2014.

\bibitem{NakagawaNOEHSK17}
S.~Nakagawa, T.~Nishide, E.~Okamoto, K.~Emura, G.~Hanaoka, Y.~Sakai, and
  A.~Kodate, ``A privacy-enhanced access log management mechanism in {SSO}
  systems from nominative signatures,'' International Journal of Applied
  Cryptography, vol.3, no.4, pp.394--406, 2017.

\bibitem{ETHYellow}
G.~Wood, ``Ethereum yellow paper ({S}hanghai version 9fde3f4-2024-09-02),''
  2024.
\newblock \url{https://ethereum.github.io/yellowpaper/paper.pdf}.

\bibitem{CramerDM00}
R.~Cramer, I.~Damg{\aa}rd, and P.D. MacKenzie, ``Efficient zero-knowledge
  proofs of knowledge without intractability assumptions,'' Public Key
  Cryptography, pp.354--373, 2000.

\bibitem{Waters05}
B.~Waters, ``Efficient identity-based encryption without random oracles,''
  {EUROCRYPT}, pp.114--127, 2005.

\bibitem{KansalDM21}
M.~Kansal, R.~Dutta, and S.~Mukhopadhyay, ``Lattice-based nominative signature
  using pseudorandom function,'' {IET} Information Security, vol.15, no.4,
  pp.317--332, 2021.

\bibitem{FerschKP16}
M.~Fersch, E.~Kiltz, and B.~Poettering, ``On the provable security of {(EC)DSA}
  signatures,'' {ACM} {CCS}, 2016.

\bibitem{GrothS22}
J.~Groth and V.~Shoup, ``On the security of {ECDSA} with additive key
  derivation and presignatures,'' {EUROCRYPT}, pp.365--396, 2022.

\end{thebibliography}

\end{document}